# Gain-assisted superluminal light propagation via incoherent pump field

M. Mahmoudi [(1)], S. Worya Rabiei [(2)], L. Safari [(1)] and M. Sahrai [(3)]

[(1)] Department of Physics, Zanjan University, P.O.Box 45195-313, Zanjan, Iran

[(2)] Department of Physics, University of Kurdistan, Sanandaj, Iran

[(3)] Research Institute for Applied Physics and Astronomy,

University of Tabriz, Tabriz, Iran

**Abstract**

We investigate the dispersion and the absorption properties of a weak probe field in a three-level $\Lambda$-type atomic system. We use just an incoherent field for controlling the group velocity of light. It is shown that the slope of dispersion changes from positive to negative just with changing the intensity of the indirect incoherent pumping field. Gain-assisted superluminal light propagation appears in this system. No laser field is used in the pumping processes.

**Keywords:** incoherent field, susceptibility, group velocity

Although the problem of propagation of a light pulse inside dispersive media has been studied for more than a century, it still attracts considerable interest. This is because of the discovery that the velocity of pulse, assumed to be the group velocity, can be either larger (superluminal light propagation) or smaller (subluminal light propagation) than the speed of light in a vacuum or even become negative inside a dispersive medium [1, 2]. Sommerfeld and Brillouin were among the early researchers studying wave propagation in linear, homogeneous, isotropic, causally dispersive media [3]. The theory of superluminal propagation has actively been developed since 1990's [4-8].

Recently, slow or even stopped group velocities and superluminal propagation of light waves have been demonstrated in specific classes of atomic and solid state materials [9-11]. The origin of these effects is the high dispersion properties of the materials. The regions of sharp normal dispersion give rise to a very slow group velocity, while the regions of strong anomalous dispersion give rise to the superluminal effects. The extended scientific discussion about superluminal light propagation can be found in review articles [12, 13].

Despite the name superluminal light, it is generally believed that no information can be sent faster than light speed $c$ in vacuum, as explained by Chiao [12] and Gauthier [14]. Thus, a group velocity faster than $c$ does not violate Einstein's principle of special relativity.

It was shown that switching from subluminal to superluminal pulse propagation can be achieved by the intensity of a coupling field [15-18], and relative phase between two weak probe fields [19]. The incoherent pumping fields can also play an important role in the controlling of the group velocity of light in dispersive media. Recently, we have studied the effects of an incoherent pumping field and the spontaneously generated coherence (SGC) on the phase control of group velocity in a three-level $\Lambda$-type system [20] and it was shown that the phase-dependent superluminal light propagation appears near an absorption peak. Korsunsky et al. presented a theory of a continuous wave light propagation in a medium of atoms with a double $\Lambda$ configuration [21]. They have shown that when the so-called multi-photon resonance condition is fulfilled, both absorptive and dispersive properties of such medium depend on the relative phase of the driving fields. Recently, the light propagation through closed-loop atomic medium beyond the multi-photon resonance condition has also been studied. It is shown that the medium response oscillating in phase with the probe field, in general, is not phase-dependent and for parameters violating the multiphoton resonance condition, inducing a closed loop interaction contour is not advantageous [22].

Wang et al have also observed the gain-assisted superluminal light propagation in an atomic three-level Cesium (Cs) vapor cell via the coherent pumping fields [9]. We have investigated the dispersion and the absorption properties of a weak probe field in a four-level atomic system by using incoherent pumping fields. It was demonstrated that the group velocity of a light pulse can be controlled by the rate of the incoherent pumping fields, however, the superluminal light propagation was accompanied by an absorption peak [23].

In this letter we investigate the dispersion and the absorption properties of a weak probe field in a simple three-level quantum system. We use just an indirect incoherent pumping field to control the group velocity of light. It is shown that in our proposed scheme the group velocity of light can be controlled by changing the indirect incoherent pumping field and the superluminal light propagation is accompanied by a gain doublet. To compare this scheme with the ones presented in our recent papers [23, 24], one should note that in the first one, despite the ability to control the group velocity of light without using coherent driving fields, the superluminal light propagation was accompanied by absorption. In the second one, a driving laser field did accompany the indirect incoherent pumping field in the switching of the slope of dispersion from positive to negative.

In the following, we present the model, basic equations, analytical expressions, and results of the susceptibility and the group index.

Let us consider a closed three-level atomic system with two well-separated lower-levels $|1\rangle$ and $|2\rangle$, and an upper level $|3\rangle$ as shown in Fig. 1(a). The population of two lower levels, $|1\rangle$ and $|2\rangle$, are pumped to the excited state $|3\rangle$, through an auxiliary state $|\tilde{3}\rangle$, by using a direct incoherent pumping field with a spectral width greater than the separation of the lower states. Under this condition, the indirect incoherent pumping process excites both transitions, $|1\rangle \leftrightarrow |3\rangle$ and $|2\rangle \leftrightarrow |3\rangle$. A weak tunable probe field of frequency $\omega_p$ is applied to both transitions $|1\rangle \leftrightarrow |3\rangle$ and $|2\rangle \leftrightarrow |3\rangle$. The spontaneous decay rates from level $|3\rangle$ to the lower levels $|1\rangle$ and $|2\rangle$ are denoted by $\gamma_1$ and $\gamma_2$, respectively.

There are three major dynamical processes occurring in the system: $(i)$ indirect incoherent pumping processes through $R_1$, $R_2$ $(ii)$ interaction with the reservoir governing the decay processes from the level $|3\rangle$ to the levels $|1\rangle$ and $|2\rangle$ $(iii)$ interaction with the weak coherent probe field. The processes are described by three interaction Hamiltonian terms $H_1$, $H_2$, and $H_3$. Including the free energy term ($H_0$), the total Hamiltonian is [25]

$$H = H_0 + H_1 + H_2 + H_3, \qquad (1)$$

where

$$H_0 = \hbar\omega_1 |1\rangle\langle 1| + \hbar\omega_2 |2\rangle\langle 2| + \hbar\omega_3 |3\rangle\langle 3|, \qquad (2.a)$$

$$H_1 = -P_1 E_p |\tilde{3}\rangle\langle 1| - P_2 E_p |\tilde{3}\rangle\langle 2| + c.c., \qquad (2.b)$$

$$H_2 = -\hbar\sum_k g_k^{(1)} e^{i(\omega_{31}-\nu_k)t} |3\rangle\langle 1|\hat{b}_k - \hbar\sum_k g_k^{(2)} e^{i(\omega_{32}-\nu_k)t} |3\rangle\langle 2|\hat{b}_k + c.c., \qquad (2.c)$$

$$H_3 = -\hbar\Omega_{p1} e^{-i\omega_p t} |3\rangle\langle 1| - \hbar\Omega_{p2} e^{-i\omega_p t} |3\rangle\langle 2| + c.c., \qquad (2.d)$$

where $\hbar\omega_i$ gives the energy of the state $|i\rangle$ ($i=1,2,3$). The state $|\tilde{3}\rangle$ shows an auxiliary state that is necessary to establish the indirect incoherent pumping from states $|1\rangle$ and $|2\rangle$ to the state $|3\rangle$. Here $\Omega_{p1} = \dfrac{\varepsilon_p \wp_{13}}{2\hbar}$ and $\Omega_{p2} = \dfrac{\varepsilon_p \wp_{23}}{2\hbar}$ are the Rabi frequencies of the weak probe field, corresponding to the transitions $|1\rangle \leftrightarrow |3\rangle$ and $|2\rangle \leftrightarrow |3\rangle$, respectively. Here $\wp_{i3}$ represents the atomic dipole moments and $\varepsilon_p$ is the amplitude of the probe field. $g_k^{(1)}$ and $g_k^{(2)}$ are the coupling constants between the $k$ th vacuum mode of frequency $\nu_k$ and the atomic transitions from level $|3\rangle$ to levels $|1\rangle$ and $|2\rangle$. $P_1$ and $P_2$ are the dipole moments of the atomic transitions corresponding to the pumpings from the levels $|1\rangle$ and $|2\rangle$ to the level $|\tilde{3}\rangle$, respectively. $E_p$ is the amplitude of the incoherent pumping field. $b_k (b_k^t)$ is the annihilation (creation) operator for the $k$ th vacuum mode with frequency $\nu_k$; $k$ here represents both the momentum and polarization of the vacuum mode.

This system can be understood in the context of a four-level system shown in Fig. 1(b). In this scheme an incoherent pumping field is applied to $|1\rangle \leftrightarrow |\tilde{3}\rangle$ and $|2\rangle \leftrightarrow |\tilde{3}\rangle$ transitions and the population is pumped from $|1\rangle$ and $|2\rangle$ to $|\tilde{3}\rangle$. The spontaneous decay rates from $|\tilde{3}\rangle$ to the levels $|3\rangle$, $|2\rangle$, and $|1\rangle$ are denoted by $\Gamma_0$, $\Gamma_2$, and $\Gamma_1$, respectively. Note that for $\Gamma_0 \gg \Gamma_1, \Gamma_2$, such a four-level system is equivalent to the proposed three-level quantum system driven by an indirect incoherent pumping field. Such a system was used to establish the lasing without inversion and enhancement of the index of refraction via interference of incoherent pumping processes [26].

The density matrix equations of motion under rotating wave approximation and in the rotating frame are:

$$\dot{\rho}_{11} = i\, \Omega_{p_1} \rho_{31} - i\, \Omega_{p_1}^* \rho_{13} + \gamma_1 \rho_{33} - R_1 \rho_{11}, \tag{3.a}$$

$$\dot{\rho}_{22} = i\, \Omega_{p_2} \rho_{32} - i\, \Omega_{p_2}^* \rho_{23} + \gamma_2 \rho_{33} - R_2 \rho_{22}, \tag{3.b}$$

$$\dot{\rho}_{21} = (i\,\omega - \gamma_{21})\rho_{21} + i\, \Omega_{p_2} \rho_{31} - i\, \Omega_{p_1}^* \rho_{23}, \tag{3.c}$$

$$\dot{\rho}_{31} = [i\,(\delta_p + \tfrac{\omega}{2}) - \gamma_{31}]\rho_{31} + i\, \Omega_{p_2}^* \rho_{21} - i\, \Omega_{p_1}^* (\rho_{33} - \rho_{11}), \tag{3.d}$$

$$\dot{\rho}_{32} = [i\,(\delta_p - \frac{\omega}{2}) - \gamma_{32})]\rho_{32} + i\,\Omega_{p_1}^* \rho_{12} - i\,\Omega_{p_2}^* (\rho_{33} - \rho_{22}), \qquad (3.e)$$

$$\rho_{11} + \rho_{22} + \rho_{33} = 1, \qquad (3.f)$$

where $\hbar\omega$ is the energy difference of the lower levels. The parameter $\delta_p = \omega_p - \frac{\omega_{31} + \omega_{32}}{2}$ measures the common detuning of probe field from the middle of the lower levels. The quantum interference due to the incoherent pumping field is dropped. We have also ignored the quantum interference due to the spontaneous emissions from the upper level. It was shown that only for nearly degenerate lower-levels, i.e. $\omega_{13} \approx \omega_{12}$, this interference becomes important, and for large separation between the lower levels it may be dropped [27, 28].

The steady state solutions for a weak probe field, i.e. $\Omega_{p1}, \Omega_{p2} \ll \gamma$, are

$$\rho_{11} = \frac{R_2 \gamma_1}{R_1 R_2 + R_1 \gamma_2 + R_2 \gamma_1}, \qquad (4.a)$$

$$\rho_{22} = \frac{R_1 \gamma_2}{R_1 R_2 + R_1 \gamma_2 + R_2 \gamma_1}, \qquad (4.b)$$

$$\rho_{33} = \frac{R_1 R_2}{R_1 R_2 + R_1 \gamma_2 + R_2 \gamma_1}, \qquad (4.c)$$

$$\rho_{31} = \Omega_{P_1}^* \frac{\rho_{33} - \rho_{11}}{(\delta_P + \frac{\omega}{2}) + i\,\gamma_{31}}, \qquad (4.d)$$

$$\rho_{32} = \Omega_{P_2}^* \frac{\rho_{33} - \rho_{22}}{(\delta_P - \frac{\omega}{2}) + i\,\gamma_{32}}, \qquad (4.e)$$

where

$$\gamma_{31} = \frac{1}{2}(\gamma_1 + \gamma_2 + R_1), \qquad \gamma_{32} = \frac{1}{2}(\gamma_1 + \gamma_2 + R_2), \qquad \gamma_{21} = \frac{1}{2}(R_1 + R_2).$$

In the following, it is assumed that the dipole matrix elements for both transitions $|1\rangle \leftrightarrow |3\rangle$ and $|2\rangle \leftrightarrow |3\rangle$ are equal, i.e. $|\vec{\wp}_{13}| = |\vec{\wp}_{23}| = \wp$, so the common Rabi-frequency is $\Omega_{p1} = \Omega_{p2} = \Omega_p = \frac{\varepsilon_p \wp}{2\hbar}.$

The response of the atomic system to the applied fields is determined by the susceptibility $\chi$, which is defined as [28]

$$\chi(\nu_p) = \alpha \frac{1}{\Omega_p}(\rho_{31} + \rho_{32}), \tag{5}$$

where $\alpha = \frac{N\wp^2}{\varepsilon_0 \hbar}$, and $N$ is the atom number density in the medium. The real and imaginary parts of $\chi$ correspond to the dispersion and the absorption of the weak probe field, respectively. For further discussion we introduce the group index $n_g = \frac{c}{v_g}$ where c is the speed of light in vacuum and the group velocity $v_g$ is given by [5, 9]

$$v_g = \frac{c}{1 + 2\pi\chi'(\nu_p) + 2\pi\nu_p \frac{\partial \chi'(\nu_p)}{\partial \nu_p}}. \tag{6}$$

The slope of the dispersion spectrum versus the probe field detuning has a major role in the calculation of the group velocity. In our notation the negative slope of dispersion corresponds to superluminal light propagation, while the positive slope shows the subluminal light propagation. Also, if $\text{Im}[\chi] < 0$, the system exhibits gain for the probe field, whereas for $\text{Im}[\chi] > 0$, the probe field is attenuated.

In the case of $R_1 = R_2 = R$ and $\gamma_1 = \gamma_2 = \gamma$ the population difference of the levels can be written as

$$\rho_{33} - \rho_{11} = \rho_{33} - \rho_{22} = \frac{R-\gamma}{R+2\gamma}. \tag{7}$$

By substituting eqs. (4.d) and (4.e) in eq. (5) the real and imaginary parts of the susceptibility are given by:

$$\text{Re}[\chi] = \alpha \left(\frac{R-\gamma}{R+2\gamma}\right)\left(\frac{(\delta_p + \omega/2)}{(\delta_p + \omega/2)^2 + \gamma_r^2} + \frac{(\delta_p - \omega/2)}{(\delta_p - \omega/2)^2 + \gamma_r^2}\right), \tag{8.a}$$

$$\text{Im}[\chi] = -\alpha \left(\frac{R-\gamma}{R+2\gamma}\right)\left(\frac{\gamma_r}{(\delta_p + \omega/2)^2 + \gamma_r^2} + \frac{\gamma_r}{(\delta_p - \omega/2)^2 + \gamma_r^2}\right), \tag{8.b}$$

where $\gamma_r = \gamma + R/2$.

The imaginary part of $\chi$ is a linear superposition of two Lorentzian functions with full width $2\gamma_r$, located at $\delta_p = \pm\frac{\omega}{2}$.

For $R = \gamma$, the populations of the levels are equal and the system is saturated, while for $R > \gamma$ the population inversion is established, and then the amplification with inversion of the probe field appears in the system.

In Fig. 2 we display the subluminal and superluminal regions via indirect incoherent pumping rate and $\omega$. The two superluminal regions are shown with dark color. For $R < \gamma$ and $\omega < 2 + R$ ($\omega < 2\gamma_r$) the superluminal light propagation is accompanied by a single absorption peak (Fig 3), while for $R > \gamma$ and $\omega > 2 + R$ ($\omega > 2\gamma_r$), it is accompanied by a gain doublet (Fig 4). Two interesting regions for our aim are the subluminal with absorption doublet and the superluminal with gain doublet. Then, the minimum value for the energy difference of the lower levels has a major role in switching the group velocity from subluminal to superluminal light propagation.

For simplicity, the Rabi-frequencies are assumed to be real and all figures are plotted in the $\gamma$ unit. Firstly, we are interested to study the effect of incoherent pumping field on the dispersion and absorption spectrum of a system with $\omega < 2\gamma$.

In Fig. 3 we display the dispersion (a, c) and absorption (b, d) of a weak probe field versus probe detuning. In Figs. 3 (a) and 3(b) two small incoherent pumping rates of $R = 0.0$ (solid) and $R = 0.8\gamma$ (dashed) are chosen and the slope of dispersion around $\Delta_p = 0$ is negative, accompanied by an absorption peak in the spectrum. In Figs. 3 (c) and 3(d) the incoherent pumping rate is increased and the slope of dispersion changes from negative to positive. An investigation on Fig. 3(b) shows that the superluminal light propagation is occurred in a considerable absorption peak.

To establish the subluminal (superluminal) light propagation with low absorption (gain), we have to consider a system in which the energy difference of the lower levels is considerable. In such a system, the absorption or gain peaks of two probe transitions are well-separated and it is possible to change the absorption doublet to the gain doublet, just by varying the indirect incoherent pumping rate.

In Fig. 4 we consider a system with larger lower levels energy difference i.e. $\omega > 3\gamma$. We show the dispersion (a, c) and absorption (b, d) of the probe field of a system with $\omega = 8\gamma$ for different indirect incoherent pumping rates. In the absence of indirect incoherent pumping field, the slope of dispersion around zero detuning is positive, corresponding to the subluminal light propagation. In the presence of a weak indirect incoherent pumping field of the rate $R = 0.8\gamma$, the height of the absorption peaks decreases and the system approaches to the saturation condition. The slope of the dispersion around zero detuning is still positive. By increasing the indirect incoherent pumping rate, i.e. $R = 1.3\gamma$ (solid) and $R = 2.3\gamma$ (dashed), the slope of dispersion switches from positive to negative, corresponding to the superluminal light propagation. Then, in a system with a larger energy difference of the lower levels, by increasing the incoherent pumping rate the slope of dispersion changes from positive to negative. The absorption doublet in the spectrum is also switched to gain doublet. Note that for much more indirect incoherent pumping rates, i.e. $R > \omega - 2\gamma$ ($2\gamma_r > \omega$), the full width of the gain dips are increased to form a single one due to the power broadening effect. Then, the negative slope of dispersion is again switched to the positive one (Fig. 2).

Fig. 5 shows the group index of a light pulse versus indirect incoherent pumping rate for various lower levels energy difference of the systems with $\omega = 1.0\gamma$ (solid), $2.0\gamma$ (dashed), $8.0\gamma$ (dotted). When $\omega = 1.0\gamma$, by increasing the indirect incoherent pumping rate the system switches from negative to positive group index at $R = \gamma$, but in negative region, the probe field experiences a strong absorption. When $\omega = 2.0\gamma$, the negative region is also accompanied by a strong absorption peak. The interesting results of our paper are obtained in the systems in which the energy difference of the lower levels is bigger than the full width of the peaks ($\omega > 2\gamma_r$). When $\omega = 8.0\gamma$ and $R < \gamma$, the system has a positive group index between the absorption lines, while for $\gamma < R < 6\gamma$ the system has a negative group index between the gain lines.

Similar results can be obtained from a V-type three-level quantum system. In Fig. 1(c) we schematically show such a system in which an indirect incoherent pumping field is applied to both $|3\rangle \leftrightarrow |1\rangle$ and $|3\rangle \to |2\rangle$ transitions. A weak probe field is also applied to both transitions. The real and imaginary parts of the susceptibility can be written as

$$\text{Re}[\chi] = \alpha \left( \frac{R-\gamma'}{2R+\gamma'} \right) \left( \frac{(\delta_p + \omega'/2)}{(\delta_p + \omega'/2)^2 + \gamma_r'^2} + \frac{(\delta_p - \omega'/2)}{(\delta_p - \omega'/2)^2 + \gamma_r'^2} \right), \qquad (9.\text{a})$$

$$\text{Im}[\chi] = -\alpha \left( \frac{R - \gamma'}{2R + \gamma'} \right) \left( \frac{\gamma'_r}{(\delta_p + \omega'/2)^2 + \gamma'^2_r} + \frac{\gamma'_r}{(\delta_p - \omega'/2)^2 + \gamma'^2_r} \right), \quad (9.b)$$

where $\gamma'_r = (\gamma' + R)/2$ and $\gamma'$ denotes the spontaneous emission rate from the upper levels to the ground level. The energy difference of the upper levels in the V-type three-level system is shown by $\hbar\omega'$.

Note that the differences of eq. (8) and eq. (9) are just in the full width of peaks (or dips) and the content of the first parentheses which are the population difference of the levels (see eq. 7). Then in the case of V-type systems, similar figures to the presented ones (Fig. 2 to Fig. 5) can be plotted.

Our analytical calculation shows that the full width and separation distance of two absorption peaks (or gain dips) are important parameters for determination of the subluminal or superluminal light pulse propagation. When the separation distance of the two peaks (dips) is less than the full width of the peaks (dips), two peaks (dips) are not distinguishable and seem to be a single one. Otherwise, two absorption peaks (gain dips) of doublet are distinguishable and the system shows the subluminal (superluminal) between the peaks (dips) of doublet. Note that for $R > \omega - 2\gamma$ the negative group index switches to the positive one. It is an important point that a magnetic field can be used to control the energy difference of the lower levels ($\hbar\omega$).

The physics of the results is interesting. The indirect incoherent pumping field pumps the population to the excited states. But for $R < \gamma$ the population inversion is not established and the system shows absorption in both $|1\rangle \leftrightarrow |3\rangle$ and $|2\rangle \leftrightarrow |3\rangle$ transitions. For $R = \gamma$, the system is saturated and both the absorption and dispersion are zero. By increasing the indirect incoherent pumping rate, i.e. $R > \gamma$, the population inversion is established and then system shows gain with inversion. Note that the redistribution of the population via indirect incoherent pumping field results in changing the coherence of the system.

In conclusion, we have controlled the dispersion and the absorption of a weak probe field in a $\Lambda$-type three-level quantum system via an indirect incoherent pumping field. We introduced the subluminal and superluminal regions for such a system and showed that by changing the rate of the indirect incoherent pumping field, the subluminal light propagation switches to the superluminal light propagation. It was demonstrated that the energy difference of lower levels has a major role to establish the gain-assisted superluminal light propagation. In addition, we showed that the similar results can be obtained in a V-type three-level system. Note that we did not use any laser field in the controlling of the optical properties of the system, but on the other hand by applying a magnetic field the energy difference of the lower levels can be controlled.

**Figure captions**

Fig. 1) Proposed level schemes. a) Three-level Λ-type system driven by an indirect incoherent pumping field (dashed) and a weak probe field (solid). b) A four-level system driven by a direct incoherent pumping field (dashed) and a weak probe field (solid). c) The equivalent V-type three-level system driven by an indirect incoherent pumping field (dashed) and a weak probe field (solid).

Fig. 2) The subluminal and superluminal regions via the frequency of the energy difference of the lower levels and the indirect incoherent pumping rate. The superluminal regions are shown with dark color.

Fig. 3) Real (a, c) and imaginary (b, d) parts of susceptibility versus probe field detuning for the parameters $\omega = \gamma$, $\gamma_1 = \gamma_2 = \gamma$, $\Omega_p = 0.01\gamma$, a) and b) $R = 0.0$ (solid), $0.8\gamma$ (dashed). c) and d) $R = 1.3\gamma$ (solid), $2.3\gamma$ (dashed).

Fig. 4) Similar to Fig.3, for $\omega = 8\gamma$.

Fig.5) The group index $(\frac{c}{v_g} - 1)$ versus the indirect incoherent pumping rate for various energy differences of the lower levels, i.e. $\omega = 1.0\gamma$ (solid), $2.0\gamma$ (dashed), $8.0\gamma$ (dotted). The common parameters are $\gamma_1 = \gamma_2 = \gamma$, $\Omega_p = 0.01\gamma$, $v_p = \frac{\gamma}{2\pi}$.

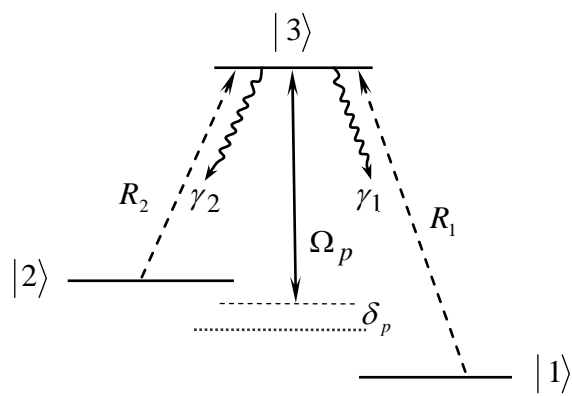

**Fig.1 (a)**

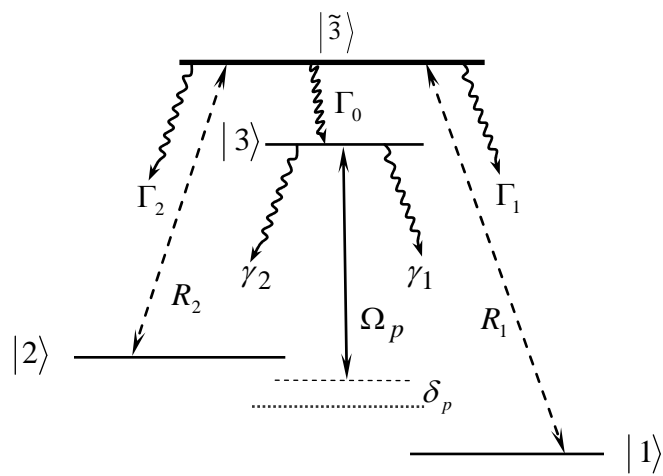

**Fig.1 (b)**

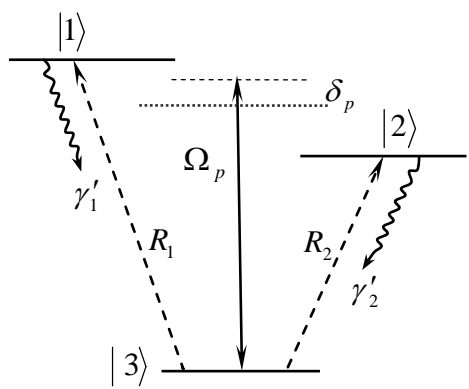

**Fig.1 (c)**

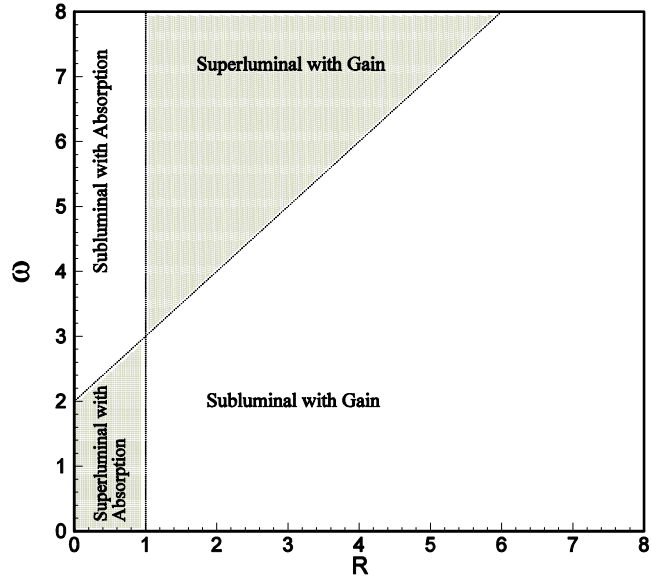

Fig. 2

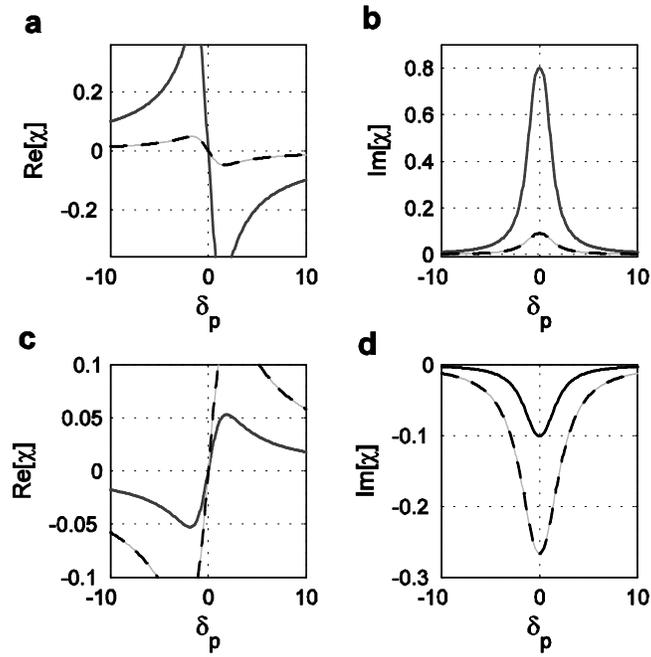

Fig. 3

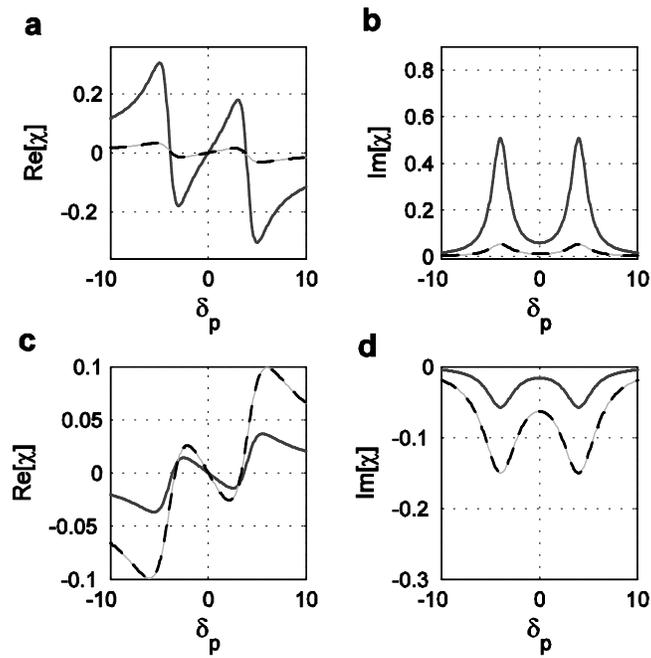

**Fig. 4**

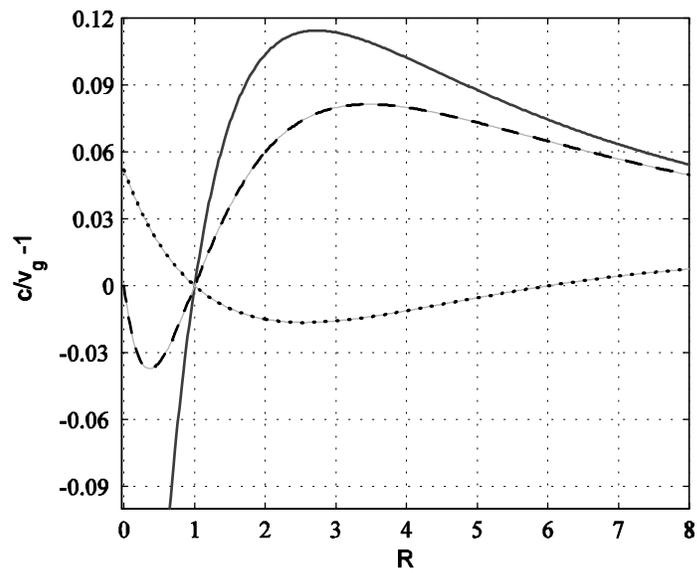

**Fig. 5**